\documentclass[amsmath,amssymb,superscriptaddress,widetext,10pt]{revtex4}
\usepackage[dvips]{graphicx}

\newcommand{\Tmelt}{\ensuremath{T_{\mathrm{M}}}}
\newcommand{\Tspin}{\ensuremath{T_{\mathrm{S}}}}

\newcommand{\lcrit}{\ensuremath{l_\mathrm{crit}}}
\newcommand{\gammaSV}{\ensuremath{\gamma_\mathrm{SV}}}
\newcommand{\gammaLV}{\ensuremath{\gamma_\mathrm{LV}}}
\newcommand{\gammaSL}{\ensuremath{\gamma_\mathrm{SL}}}

\begin{document}
\title{The NaCl\,(100) Surface: Why Does it Not Melt?
  \footnote{to appear in:
  \emph{``Highlights in the quantum theory of condensed matter''}
  (in honor of Mario Tosi), Edizioni della Scuola 
  Normale Superiore, Pisa, Italy (2005).}
}

\author{T. Zykova-Timan}
\affiliation{International School for Advanced Studies (SISSA),
via Beirut 2, 34014 Trieste, Italy}
\affiliation{\it INFM Democritos National Simulation Center,
via Beirut 2, 34014 Trieste, Italy}
\author{U. Tartaglino}
\affiliation{International School for Advanced Studies (SISSA),
via Beirut 2, 34014 Trieste, Italy}
\affiliation{\it INFM Democritos National Simulation Center,
via Beirut 2, 34014 Trieste, Italy}
\affiliation{IFF, FZ-J\"ulich, 52425 J\"ulich, Germany}
\author{D. Ceresoli}
\affiliation{International School for Advanced Studies (SISSA),
via Beirut 2, 34014 Trieste, Italy}
\affiliation{\it INFM Democritos National Simulation Center,
via Beirut 2, 34014 Trieste, Italy}
\author{E. Tosatti\footnote{Corresponding author: tosatti@sissa.it}}
\affiliation{International School for Advanced Studies (SISSA),
via Beirut 2, 34014 Trieste, Italy}
\affiliation{\it INFM Democritos National Simulation Center,
via Beirut 2, 34014 Trieste, Italy}
\affiliation{International Center for Theoretical Physics (ICTP),
Strada Costiera 11, 34014, Trieste, Italy}

\begin{abstract}
The high temperature surface properties of alkali halide crystals are very 
unusual. Through molecular dynamics simulations based on Tosi-Fumi potentials, 
we predict that crystalline NaCl\,(100) should remain stable without any 
precursor signals of melting up to and even above the bulk melting point 
$T_m$. In a metastable state, it should even be possible to overheat 
NaCl\,(100) by at least 50\,K.  The reasons leading to this lack of surface 
self-wetting are investigated. We will briefly discuss the results of calculations
of the solid-vapor and liquid-vapor interface free energies, showing that
the former is unusually low and the latter unusually high, and explaining why.
Due to that the mutual interaction among solid-liquid and liquid-vapor 
interfaces, otherwise unknown, must be strongly attractive at short distance, 
leading to the collapse of any liquid film  attempting to nucleate at the 
solid surface. This scenario naturally explains the large incomplete wetting 
angle of a drop of melt on NaCl\,(100).  
\end{abstract}

\maketitle

\section{Introduction}\label{sec:intro}
Interest is increasing toward 
adhesion, the structure and physics of solid-liquid
interfaces, and the structure of liquid surfaces, particularly of complex and molecular
systems. In order to gain more insight into these problems, there is a strong need for
good case studies, to use as well-understood starting points.

One of the easiest starting points may be to study the relationship and contact of a
liquid with its own solid, a clear situation where there will be no ambiguity of physical
description, no uncertainty in chemical composition, no segregation phenomena,
all of them complications present in the study of adhesion between different substances.

Adhesion of a liquid onto the surface of its own solid usually materializes
spontaneously with temperature. Most solid surfaces are known to wet themselves
spontaneously with an atomically
thin film of melt, when their temperature $T$ is brought close enough to the
melting point $T_m$ of the bulk solid. The phenomenon whereby the thickness
$l(T)$ of the liquid film diverges continuously and critically
as $T \to\, T_m$\, is commonly referred to as (complete)
surface melting\cite{vanderveen88,physrep}.

There are actually a number of exceptions to this behavior. Some solid surfaces 
in particular remain fully crystalline as
$T \to T_m$. This {\em surface non-melting} phenomenon, originally discovered in
molecular dynamics simulations of Au(111)\cite{carnevali} and independently observed
experimentally in Pb(111)\cite{frenken87}, is known for the close-packed face of
other metals too, such as Al(111)\cite{Al111}.

Here we are concerned with the surface of alkali halides, crystals well known
for their unusually stable neutral (100) faces. Addressing a long time ago the NaCl\,(100) 
surface, bubble experiments by Mutaftschiev and coworkers 
revealed incomplete wetting of the solid surfaces by their own
melt\cite{mutaftschiev75,mutaftschiev97}, moreover with an extraordinarily 
large partial wetting angle of $48^{\circ}$. This kind of incomplete wetting, as
is physically clear, and as was demonstrated on metals
surfaces\cite{ditolla95,physrep}, is associated with non-melting of the
crystal surface. 

In this paper we will review our recent theory work, where we showed by direct 
simulation the surface non-melting of NaCl\,(100)\cite{tanya,tanyalong}. 
The reasons leading to this kind of surface non-melting are investigated. 
First, we will show the solid surface free energy calculated by thermodynamic 
integration, and see that at high temperature it drops
due to a larger anharmonicity than in most other solid surfaces. Next,
we will examine the surface tension of the liquid NaCl surface, 
we will find unusually large, owing to a surface entropy deficit, connected
with the local surface short range molecular order. The
solid-liquid interface free energy will finally be argued to be large,
due to a 26$\%$ density difference.
We will also discuss qualitatively -- were hypothetically the solid-vapor 
interface to split into a pair of solid-liquid and  liquid-vapor interfaces,
with a thickness $l$ of liquid in between -- the interaction free energy 
$V(l)$ expected between them. This interaction is here strongly attractive
at short range,
leading to the collapse of any liquid film attempting to nucleate at the
solid-vapor interface, and causing surface non-melting. 

All quantitative results reviewed here were derived by means of 
molecular dynamics simulations, carried out extensively for NaCl\,(100) 
slabs. These simulations, as will be detailed below, are entirely based on 
interatomic potentials that Mario Tosi refined and published, together
with Fausto Fumi, just over 40 years ago\cite{tosi64}. It is a fitting
tribute to Mario's scientific perseverance, thoroughness, and general 
dependability, that these potentials still turn out to be so incredibly 
accurate, even well outside the range of temperatures
for which they were constructed and tested, such a long time ago.

\section{Simulations with Tosi-Fumi Potentials}\label{sec:simulations}
The high temperature properties of solid NaCl bulk and NaCl\,(100) slabs 
were studied by classical molecular dynamics (MD) simulations. NaCl was 
described with the potential of Tosi and Fumi\cite{tosi64} who accurately 
parametrized a Born-Mayer-Huggins form. 
The Coulomb long-range interactions were treated by the standard three
dimensional (3D) Ewald method applied to a geometry consisting
of infinitely repeated identical crystal slabs. Simulated systems generally 
comprised about 2000--5000 molecular units, with a time-step of 1\,fs,
and a typical simulation time of 200\,ps. Long-range forces
severely limit size and time in these simulations by comparison with
the order-of-magnitude larger sizes and longer times typically affordable for 
systems with short range forces\cite{ditolla95,ditollabook}.  We took
explicit care to ensure that all our results are not crucially affected by 
small sizes, and that full equilibration was achieved in all cases. 
We checked that 80\,\AA\ of vacuum between repeated slabs are sufficient
to prevent the interaction of a liquid slab with its own
replicas\cite{tanyalong}.
Calculations were done at constant cell size and with periodic boundary
conditions.  Thermal expansion was taken care of by readjusting the $(x,y)$ 
size of the cell at each temperature so as to cancel the $(x,y)$ stress in 
the bulk solid. The theoretical thermal expansion was 4.05$\times 10^{-5}$\,K$^{-1}$, 
compared with 3.83$\times 10^{-5}$\,K$^{-1}$ in experiment. The theoretical 
bulk melting temperature \Tmelt\ of NaCl was calculated by two phase coexistence 
to be $1066\pm20$\,K. Remarkably, this Tosi-Fumi melting temperature
is extremely close to the experimental melting temperature of 1073.8\,K.  
The volume expansion at melting is about $(27\pm 2)\%$, also in excellent
agreement with the experimental value of $26\%$. 

Tab.~\ref{tab1} lists some of the calculated thermodynamical quantities
at high temperature, close to the melting point.
These results provide an 
independent confirmation of the outstanding quality of the Tosi-Fumi 
description of thermodynamics of NaCl, even at very high temperatures
and as we shall see even at surfaces, where it was by no way guaranteed.

\begin{table}\begin{center}
  \begin{tabular}{ccc}
  \hline\hline
  & Simulation & Experiment \\
  \hline
  \Tmelt (K) & 1066$\pm$20 & 1074 \\
  $\Delta V$ & 27\% & 26\% \\
  L (eV/molecule) & 0.29 & 0.29 \\
  $\Delta S_m$ ($k_B$) & 6.32 & 6.38 \\
  dP/dT (kbar/K) & 0.0311 & 0.0357 \\
  \hline
  RMDS (ave.) (\AA) & 0.60 & 0.49 \\
  $\delta$ & 20--24\% & 17--20\%~\cite{lindemann} \\
  \hline\hline
  \end{tabular}
  \caption{High temperature properties of NaCl. 
  \Tmelt is the melting temperature; $\Delta V$ is the volume jump at the
  melting point; L is the latent heat of melting; $\Delta S_m$ is the entropy
  variation at the melting point; dP/dT is the resulting Clausius-Clapeyron ratio
  at the melting point.
  RMSD is the averaged root mean square displacement of atoms at the 
  melting point; $\delta$ is the RMSD over the Na--Cl distance, for the 
  Lindemann melting criterion.}
  \label{tab1}
\end{center}\end{table}

\section{Surface Free Energies and Non-melting of NaCl(100)}\label{results}
High temperature simulations of crystalline NaCl(100) slabs directly
showed the full stability of the dry, solid surface up to \Tmelt. Moreover, a well
pronounced metastability of the slab solid faces above the melting
point \Tmelt\ indicated a clear surface non-melting behavior.
We found that a much higher (``surface spinodal'') temperature 
$\Tspin \approx \Tmelt + 150\,$K needs to be reached before the
crystalline NaCl\,(100) surface spontaneously melts. $\Tspin - \Tmelt$ thus
represents the maximum ideal overheating that a defect free NaCl\,(100)
surface can theoretically sustain without becoming spontaneously unstable against
melting. This large overheating is quite similar in magnitude to that predicted for, e.g.,
Au(111) and Al\,(111)\cite{ditolla95,ditollabook}.

At any temperature between \Tmelt\ and \Tspin, bulk melting can only 
originate through nucleation. Even though nucleation 
of the melt is in reality likely to proceed from a localized surface droplet
or defect,
it is instructive to consider a very idealized nucleus consisting of a uniform liquid film 
of thickness $l$. As a function of temperature, there will be a critical
nucleation thickness \lcrit\ decreasing from $\infty$ to zero between 
\Tmelt\ and \Tspin\cite{carnevali,ditollabook}.  The free energy difference 
per unit area between a surface with a liquid film of thickness $l$ 
and the same surface in its full crystalline state is 
\begin{equation}  \label{freeenergy}
G(l) = - \rho\lambda l \left( \frac{T}{\Tmelt} -1 \right)
       + ( \gammaSL + \gammaLV - \gammaSV )
       + V(l)
\end{equation}
The first term is the gain due to the melting of the solid
at $T>\Tmelt$. Here $\lambda$ is the latent heat per unit mass and 
$\rho$ is the liquid mass density.  The second term
$\Delta\gamma_{\infty}\equiv ( \gammaSL + \gammaLV - \gammaSV )$,
is the free energy imbalance caused by replacing the SV interface with
the SL+LV pair of interfaces, supposed to be non-interacting. The last 
term $V(l)$ is an interface interaction, representing the correction to 
$\gammaSL+\gammaLV$ when the  two interfaces are at close distance. This 
definition implies $V(+\infty)=0$ and $V(0)=-\Delta\gamma_{\infty}$. 
At very large distance the interaction disappears. The solid-vapor crystal 
surface is instead recovered when the SL and LV interfaces collapse, and the 
liquid film disappears at $l=0$. The non-melting condition $\gammaSL+\gammaLV > \gammaSV$,
or $\Delta\gamma_{\infty}>0$ implies that here the interaction 
$V(l)$ is mainly attractive.

This formulation indicates three possible origins for non-melting: an 
exceptionally low free energy \gammaSV\ of the solid surface; an unusually 
large free energy \gammaSL\ of the solid-liquid interface; a relatively high 
surface tension \gammaLV\ of liquid NaCl.
As detailed elsewhere\cite{tanya,tanyalong}, all three mechanisms are
actually relevant to NaCl(100). 

\begin{figure}
  \begin{center}
  \includegraphics[width=0.8\textwidth]{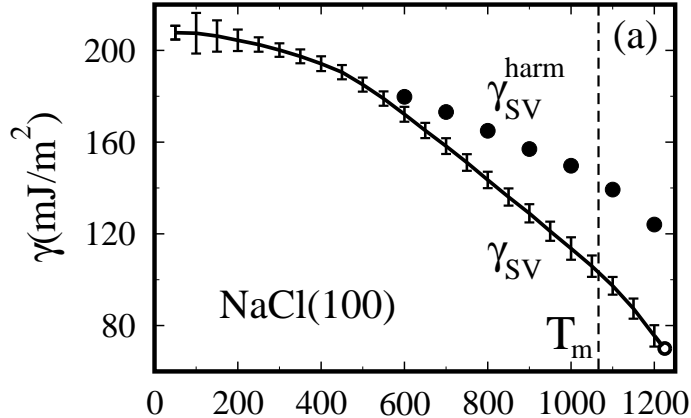}
  \caption{\label{thermo2a}
   The solid surface free energy of NaCl(100) calculated from thermodynamic
   integration (circles: effective harmonic approximation).}
\end{center}
\end{figure}

The solid-vapor interface free energy at the melting point
was calculated through standard thermodynamic integration, using 
\begin{equation}\label{eq:integration}
   \left( \frac{\partial(F/T)}{\partial(1/T)} \right)_{N, V} = E,
\end{equation}
where $E$ is the surface internal energy, extracted from simulation of the
crystalline NaCl(100) slab and the corresponding bulk
at increasing T from 50\,K to 1200\,K.
The surface free energy in Fig.\,\ref{thermo2a} shows a large drop at high temperatures, 
with an increasing deviation from an effective harmonic behavior above 600\,K, 
indicating very strong surface anharmonicity in this regime. The main
source of this anharmonicity is connected with large root
mean square thermal fluctuations of the surface Cl and Na ions above
20\% of the Na--Cl distance, largely exceeding the canonical Lindemann 
values.~\cite{tanyalong}

\begin{figure}
  \begin{center}
  \includegraphics[width=0.8\textwidth]{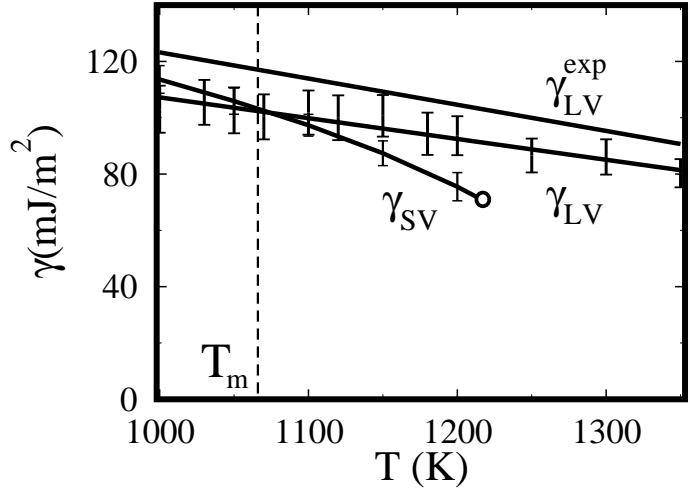}
  \caption{\label{g_fict1a}
   The calculated liquid surface free energy calculated.
   The NaCl (100) surface free energy is also shown in the temperature
   range from 1000 K to 1250 K.}
\end{center}
\end{figure}

The liquid-vapor free energy, equal to the liquid surface tension, was
evaluated from simulations of liquid NaCl slabs via the standard Kirkwood-Buff
formula\cite{tosibook}.
The first thing we note in the result, shown in Fig.\,\ref{g_fict1a}
is that right at $T_m$ the solid and liquid surface
free energies are essentially identical, 103$\pm$4 and 104$\pm$8\,mJ/m$^2$  respectively.
This is very unusual, and implies directly surface non-melting, because clearly 
$\Delta\gamma_{\infty}\equiv ( \gammaSL + \gammaLV - \gammaSV ) > 0$. In fact,
even though we did not calculate $\gammaSL$, this interface free energy has
no reason to be very small, owing to the large solid-liquid density 
difference. We independently estimated a lower bound for $\gammaSL$ to be 
36$\pm$6\,mJ/m$^2$\cite{tanyalong}.

\begin{figure}
  \begin{center}
  \includegraphics[width=0.8\textwidth]{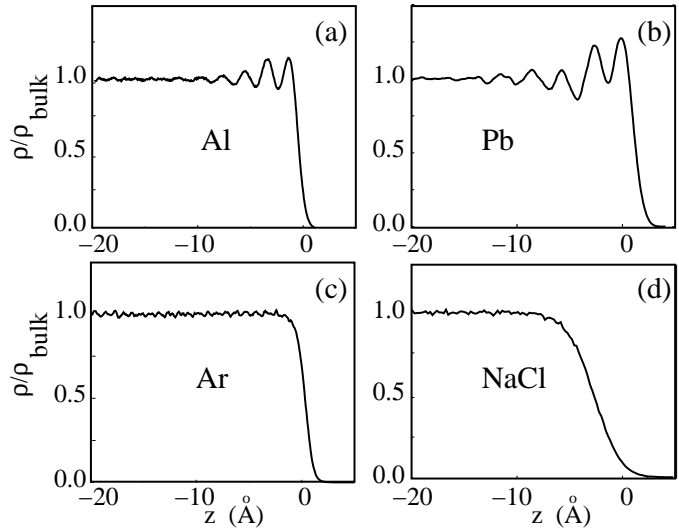}
  \caption{\label{profiles} Surface density profile of liquid NaCl, compared
with that simulated for liquid Ar, and for two liquid metals. Note that NaCl
does not show layering as the metals do, and has an even smoother profile than Ar}
  \end{center}
\end{figure}

The question that remains to be explained is therefore the physical reason 
why the liquid surface tension is so relatively high. The liquid surface density
profile, in particular, is very smooth, with none of the layering phenomena
displayed by the metal surfaces (see Fig.\,\ref{profiles})

An important clue is provided by surface entropy (per unit area)
\mbox{$S_{\mathrm{surf}} = - d\gamma/dT$}.
The temperature dependence of two surface free energies of 
Fig.\,\ref{g_fict1a} shows a factor 2.6 {\em lower} surface entropy $S_{LV}$ 
of the liquid surface compared with that of the 
solid surface.
This liquid surface entropy deficit (SED) strongly suggests 
some underlying surface short range order. Short range order can in turn 
also explain why the surface tension is here as high as
the solid surface free energy. 
The surface profile indicates that the order is clearly not layering: so what is it instead?

The answer we find is that charge order, already very important in
bulk, plays a newer and enhanced role at the molecular liquid surface. If surface
thermal fluctuations are indeed very large, we find them revealingly
{\em correlated}. For a Na$^+$ ion
 that instantaneously moves e.g., out of the surface,
there is at least one accompanying Cl$^-$, also moving out; and vice versa.
So on one hand the large fluctuations smear the average liquid vapor
density profile, bridging very gently between the liquid and essentially zero
in the vapor, (Fig.\,\ref{profiles}). On the other hand
the two body correlations, described e.g., by the the Na-Cl pair correlation
function, or by its integral, the ion coordination number $N$,
do not vanish identically in the vapor, but tend to a typical value corresponding
to the NaCl molecule, (plus in fact a large concentration of dimers, Na$_2$Cl$_2$).
The lack of freedom implied by the incipient molecular bond explains the
entropy deficit, and the consequent large surface tension of the molten salt
surface.
To confirm if this is true, we recalculated  $\gammaSL$ modifying
the forces in the Kirkwood-Buff formula by removing all contributions
from surface ions whose coordination is between zero and 1.3, the average
NaCl vapor value, which amounts to suppress the molecular order at the
liquid-vapor interface~\cite{tanya}.
This construction, meant to provide a qualitative estimate 
of where would the surface tension drop if surface molecular order were
absent, gives a surface tension of about 50\,mJ/m$^2$\cite{tanya}. With a
surface tension this low, surface non-melting would in fact disappear,
and wetting of the solid surface by the molten salt would be complete.
Hence the high surface tension of liquid NaCl can indeed be ascribed to surface molecular
short range order, ultimately due to charge neutrality.
This result confirms an early surmise by Goodisman and
Pastor\cite{goodisman}.

\section{Interface Interaction}\label{model}
In this short speculative section we further rationalize the results 
above within the phenomenological framework of Eq.(1), where besides 
the bare interface free energies just calculated, an 
interaction $V(l)$ appears. We will not present a calculation of $V(l)$, 
but simply discuss it on physical grounds, in the light of our new microscopic
understanding gained through simulations, and calculations of interface
free energies just reviewed\cite{tanya}. 
The definition of interface interaction $V(l)$ given earlier
implied $V(+\infty)=0$ and $V(0)=-\Delta\gamma_{\infty}$.
At very large distance the interaction disappears. At the opposite
limit, when the SL and LV interfaces approach each other and merge at 
very close quarters, they will eventually yield the SV interface upon 
their collapse, when the liquid film disappears altogether at $l=0$. 
The non-melting condition $\gammaSL+\gammaLV > \gammaSV$, or 
$\Delta\gamma_{\infty}>0$ implies here that the interaction $V(l)$ is 
attractive at short range.

In non-melting metal surfaces, a source of finite-range attraction was described
as the result of a constructive interference between two equal-period
damped density oscillations, one entering the liquid film from the solid side,
the other, due to surface layering, from the vacuum side. Here, one of the
two oscillations, namely that on the vacuum side, is missing, because
there is no layering at the molten salt surface. At large distance the
main interaction between the SL and the LV NaCl interfaces will essentially be
due to electrostatic forces and to dispersion forces. The latter in particular 
give rise to an additional long-range interface interaction $V_{\mathrm{dis}}(l) = H\,l^{-2}$ 
which is dominant at large distances. Here $H$ is the Hamaker
constant, that can be estimated through the formula 
$H = (\pi/12)\, C_{6} (n_{s}-n_{l})(n_{l}-n_{v}) = 0.00119\,\mbox{eV}$,
where $C_6=72.5\,\mbox{eV\,\AA}^6$ is the coefficient of the
Lennard-Jones interaction between chlorine ions, $n_{s}$, $n_{l}$ and
$n_{v}$ are the number of Cl$^{-}$ ions per unit volume respectively in
the solid, liquid and vapor phases\cite{israelachvili}. Since the liquid 
density is only about 79\% that of the solid, this constant is positive 
which implies a long range repulsion of the SL and LV interfaces. A simple 
estimate indicates however that $V_{\mathrm{dis}} < H/a^2 = 0.51\,\mbox{mJ/m$^2$}$,
a value that makes it irrelevant in practice.

Therefore we expect the effective interaction $V(l)$ to be very weak
in NaCl, everywhere except very close to zero range, $l \approx a$. Here it
will suddenly turn strongly attractive, $V(0)\approx -\Delta\gamma_{\infty}$. 
The physics of this short range attraction has already been described,
because it amounts to the free energy gained by replacing the two costly
LV and SL interfaces, with the single and less costly SV interface.

\section{Conclusions}\label{conclusion}
Summarizing, the NaCl\,(100) surface is predicted to show non-melting and 
to sustain overheating up to a theoretical maximum of about 150\,K above the bulk melting point.
The thermodynamics of surface non-melting in alkali halides is shown to
differ from that of metal surfaces, e.g.\ Al\,(111), Pb\,(111) or Au\,(111).
Unlike metals, non-melting in alkali halides is not connected with liquid layering,
but to molecular short range order raising the liquid surface tension, as well as
to strong anharmonicty that lowers the free energy of the solid surface.
It is argued moreover that the thermodynamical SL-LV interface interaction should
consist mainly of a strong short-range attraction. Fresh microscopic experimental 
work, absent so far, is called for to check these predictions on the high
temperature behavior of NaCl(100) and other alkali halide surfaces.

\section*{Acknowledgments}\label{acknow}
We wrote this paper to honor Mario Tosi on his 72th birthday. The high
accuracy and dependability which we found for his potentials is a tribute 
to his long-lasting work. This project was sponsored by the Italian Ministry 
of University and Research, through COFIN2003, COFIN2004 and FIRB RBAU01LX5H; 
and by INFM, through the``Iniziativa Trasversale Calcolo Parallelo''.
Calculations were performed on the IBM-SP4 at CINECA, Casalecchio (Bologna).
We are grateful to E. A. Jagla for his help and discussions about NaCl(100).



\begin{thebibliography}{99}
\bibitem{vanderveen88}
  J. F. van der Veen, B. Pluis and A. W. Denier van der Gon,
  in: R. Vanselow, R. F. Howe (Eds.),
  ``Chemistry and Physics of Solid Surfaces VII'',
  Springer, Heidelberg, 1988.

\bibitem{physrep}
  U. Tartaglino, T. Zykova-Timan, F. Ercolessi and E. Tosatti, 
  Phys. Reps. \textbf{411}, 291 (2005).

\bibitem{carnevali}
  P. Carnevali, F. Ercolessi, and  E. Tosatti, Phys. Rev. B
  \textbf{36} (1987), 6701.

\bibitem{frenken87}
  B. Pluis, A. W. Denier van der Gon, J. W. M. Frenken
  and J. F. van der Veen, Phys. Rev. Lett. \textbf{59} (1987), 2678.

\bibitem{Al111}
  A. W. Denier van der Gon, R. J. Smith, J. M. Gay, D. J. O'connor
  and J. F. van der Veen,
  Surf. Sci.  227, Issues 1-2 , 1 March 1990, Pages 143-149

\bibitem{mutaftschiev75}
  G. Grange and B. Mutaftschiev, Surf.Sci. \textbf{47}, 723 (1975).

\bibitem{mutaftschiev97}
  L. Komunjer, D. Clausse and B. Mutaftschiev,
  J. Cryst. Growth \textbf{182}, 205 (1997).

\bibitem{ditolla95}
  F. D. Di Tolla, F. Ercolessi and E. Tosatti,
  Phys. Rev. Lett. \textbf{74}, 3201 (1995).

\bibitem{tanya}
  T. Zykova-Timan, D. Ceresoli, U. Tartaglino and E. Tosatti,
  Phys. Rev. Lett. \textbf{94}, 176105 (2005).

\bibitem{tanyalong}
  T. Zykova-Timan, U. Tartaglino, D. Ceresoli and E. Tosatti,
  J. Chem. Phys. \textbf{123} (2005).

\bibitem{lindemann}
  M. A. Viswamitra and K. Jayalakshmi, Acta Crystllogr. A \textbf{28}, S189, 1972. 

\bibitem{tosi64}
  M. P. Tosi and F. G. Fumi, J. Phys. Chem. Solids \textbf{25}, 45 (1964).

\bibitem{ditollabook}
  F. di Tolla, E. Tosatti, and F. Ercolessi, 
  Interplay of melting, wetting, overheating and faceting on metal
  surfaces: theory and simulation,
  in: K. Binder, G. Ciccotti (Eds.),
  ``Monte Carlo and Molecular Dynamics of Condensed Matter Systems'',
  Italian Physical Society, Bologna, 1996, p.~345

\bibitem{tosibook} 
  N. H. March and M. P. Tosi, \emph{Atomic dynamics in liquids} (Macmillan, London, 1976).
 
\bibitem{zykova03} 
  T. Zykova-Timan, U. Tartaglino, D. Ceresoli, W. Sekkal-Zaoui and E. Tosatti,
  \emph{Surf. Sci.} \textbf{566-568} (2004) 794

\bibitem{goodisman}
  R. W. Pastor and J. Goodisman, 
  J. Chem. Phys. \textbf{68}, 3654 (1978)

\bibitem{israelachvili}
  J. N. Israelachvili, ``Intermolecular and Surface Forces'', p.~177,
  Academic Press, San Diego, 2nd edition, 1991.

\end{thebibliography}
\end{document}